\documentclass[prd,nofootinbib,showpacs,superscriptaddress,preprint]{revtex4}
\usepackage[T1]{fontenc}
\usepackage{amsmath,amssymb}
\usepackage{epsfig}
\usepackage{dcolumn}
\usepackage{mathalfa}
\usepackage[usenames]{color}
\usepackage{slashed}
\usepackage[colorlinks,citecolor=blue]{hyperref}
\usepackage{pdfpages}
\usepackage{float}
\usepackage{graphicx}
\usepackage{xcolor}
\usepackage{array}
\usepackage{multirow}
\usepackage[font=footnotesize]{caption}
\usepackage{pifont}
\usepackage{multirow}
\usepackage{tabularx}
\def\bea{\begin{eqnarray}}
	\def\eea{\end{eqnarray}}

\begin{document}
		
		\title{Constraints on 1-0 texture through neutrino phenomenology and dark matter in minimal inverse seesaw}
		\author{Jotin Gogoi}
		\email{jotingo@tezu.ernet.in}
		\author{Mrinal Kumar Das}
		\email{mkdas@tezu.ernet.in}
		\affiliation{Department of Physics, Tezpur University, Tezpur-784028, India}
		
		\begin{abstract}
			
			\noindent In this work we have realized texture zero structures of neutrino mass matrix through our study of neutrino phenomenology and dark matter. For analysing these processes, we have constructed a model in minimal inverse seesaw, ISS(2,3) by using $A_4$ discrete symmetry. The particle content of ISS(2.3) has been augmented by a scalar triplet $\eta=(\eta_1,\eta_2,\eta_3)$. The probable dark matter candidates for this model are the neutral components of $\eta$. The three mass matices of ISS(2,3), $M_D$, $M_{NS}$ and $M_S$ contribute to the structure of light neutrino mass matrix $m_\nu$. Here we try to examine the impact on texture structures of $m_\nu$ due to different possible 2-0 structures of $M_D$. To examine further possible contraints, we have evaluated the neutrino parameters and calculated relic density of dark matter for the favourable cases. From our analysis we find that out of the fifteen possible 2-0 structures, only two of them ($M_{D3}$ and $M_{D6}$) successfully generates all the mixing angles in the allowed ranges.

			\vspace{3cm}
		\end{abstract}

		\maketitle

	\section{Introduction}	
    In particle physics the most successful model till date is the Standard Model. It elegantly includes all the known fundamental particles and describes the relevant interactions through the gauge groups $SU(3)_C\otimes SU(2)_L\otimes U(1)_Y$. The discovery of Higgs boson in 2012 further enhanced the credibility of the model, thereby gaining a significant milestone in understanding the building blocks of nature \cite{doi:10.1063/1.4727988}. Though the Standard Model was a huge success, there are many relevant phenomena present in the universe which are yet to be explained and understood properly. These challenges come from phenomena that are observed in cosmology, astrophysics etc. Dark matter, baryon asymmetry of the universe (BAU), dark energy are some of the most common, yet unsolved mysteries of the universe. This calls for an entire new domain of physics, extending beyond the Standard Model which can provide a proper explanation to these challenging issues of the universe.
	
	Neutrino oscillations require neutrinos to be massive. The change of flavor from one generation to another is possible only when neutrinos are massive and there is mixing between them. These observations were further confirmed by the findings in experiments such as Super-Kamiokande \cite{KamLAND:2008dgz,Super-Kamiokande:2001bfk} and Sudbury Neutrino Observatories \cite{SNO:2002hgz}. Due to the absence of right-handed neutrinos the Standard Model fails to provide a proper explanation about the mass of neutrinos. This leads to the development of new alternate mechanisms which could address the issue of neutrino masses. The answers to this quest came in the form of Seesaw mechanisms, which is an extension of Standard Model by adding right-handed neutrinos. Depending on the nature of the extra particles, these are divided into Type-I \cite{Chang:1985en,Mohapatra:1987fs}, Type-II \cite{Rodejohann:2004qh}, Type-III \cite{King:2014nza,Arhrib:2009mz}, inverse seesaw \cite{Mohapatra:2005wg,Zhang:2021olk,Deppisch:2004fa}, radiative seesaw \cite{Babu:1988qv} etc. These mechanisms have been studied in great detail in ample literatures. Due to certain advantages inverse seesaw is quite popular among the model building community. One of such advantages is its capability to lower the mass scale of the right-handed neutrinos to TeV, thereby increasing the probability of detection.
	
	Another important puzzle of the universe is the presence of dark matter (DM). Several observations in astrophysics and cosmology point to the existence of a non-luminous, unknown, non-baryonic dark matter which constitutes around 26$\%$ of the universe \cite{Murayama:2007ek,Kolb:1990vq}. This amounts to roughly more than five times of the luminous matter. There are several evidences that support the presence of dark matter. Some of these prominent evidences are the  galaxy rotation curves \cite{Rubin:1970zza}, observations from the Bullet clusters \cite{Clowe:2006eq}, galaxy cluster observations made by Zwicky \cite{Zwicky:1933gu} and cosmological data from the Planck collaboration \cite{Planck:2019nip}. Though the existence of dark matter is an established fact, its nature remains elusive. In different works the particle and its nature are different i.e. fermion, scalar, vector, axions etc. The author in \cite{Taoso:2007qk} has beautifully pointed out the criteria that a particle must have in order to be a suitable dark matter candidate. Moreover, based on structure formation dark matter is primarily divided into three categories: hot dark matter, warm dark matter and cold dark matter. At present the dark matter abundance/relic density found in the universe is \cite{Planck:2018vyg}:
	\begin{equation}
		\Omega h^2=0.1199\pm 0.0027
	\end{equation}
	
	This work is based on texture zeros of neutrino mass matrix. There are plenty of works present in literature which dicuss the implications of texture zeros in different neutrino mass models. As mentioned in \cite{Ludl:2014axa,Frampton:2002yf,Xing:2002ta,Singh:2016qcf}, only one-zero (1-0) and two-zero (2-0) textures are compatible with the latest data of oscillation parameters. In this work, we have extended the minimal inverse seesaw with a scalar triplet $\eta=(\eta_1,\eta_2,\eta_3)$. One of these $\eta$'s after symmetry breaking acquire VEV while the other two remain neutral. These neutral components are the dark matter candidates of the model. The three mass matrices present in this model i.e. $M_D$, $M_{NS}$ and $M_S$ contribute to the formation of light neutrino mass matrix. The structures of these matrices significantly determine the zero structures of neutrino mass matrix. We try to study how the texture conditions are affected by these matrices. For this purpose, we have fixed the structures of $M_{NS}$ and $M_S$. After this we vary the Dirac mass matrix $(M_D)$ by considering two of its elements to be simultaneously zero. We find that out of the fifteen possible 2-0 textures of $M_D$ only six of them are able to produce 1-0 textures of the light neutrino matrix. The symmetry group used in this work is the non-abelian discrete group $A_4$. We have also used five flavons in our model. Finally we construct the lagrangian of the model and then find the $3\times3$ mass matrix for the active neutrinos. After this we have calculated the values of the model parameters and then studied neutrino phenomenology and dark matter.
	
	In this paragraph we give a brief outline of the article. Section \ref{secII} contains a brief introduction about inverse seesaw. In section \ref{secIII} we have discussed the model in detail. Here we illustrate the charge assignments of different particles, construction of the lagrangian and corresponding mass matrices of the model. In section \ref{secIV} we have discussed the texture structures for this model. In section \ref{secVI} we have discussed the phenomena of dark matter. Section \ref{secVII} contains the results of our work. Here we have discussed in detail about the results and findings from the model. Finally in section \ref{secVIII} we conclude by giving a summary of our work.

	\section{Inverse Seesaw}{\label{secII}}
	In the field of model building there are a number of frameworks/mechanisms that people adopt to study neutrino phenomenology and also physics beyond the Standard Model. Inverse seesaw is one of the important mechanisms among them. Just like the other popular mechanisms, it is also an extension of the Standard Model with the addition of a few extra particles \cite{Park:2010by,Khalil:2010iu,Ma:2009kh}. These additional particles are the right-handed neutrinos and sterile fermions. Though the number of these fermions is not fixed,  in the conventional inverse seesaw there are three right-handed neutrinos and also the sterile fermions are taken to be three in number. This takes the total number of neutral particles in the inverse seesaw to nine. The presence of sterile neutrinos facilitate to bring down the mass scale of right-handed neutrinos to TeV. By doing so it enhances the possibility of detecting these particles in the ongoing and future experiments. Based on the particles the Yukawa lagrangian  for the neutrino sector in the basis $n_L=(\nu_{L,\alpha},N^c_{R,i},S_j)^T$ is written as \cite{Dev:2009aw}:
	\begin{equation}
	\mathcal{L}=-\frac{1}{2}n^T_LCMn_L+h.c.
	\label{secIIeq1}
	\end{equation}
	The interactions between left-handed ($\nu_L$) and right-handed ($N_R$) neutrinos give rise to Dirac mass term $M_D$. Whereas the interactions that occur between $N_R$ and sterile fermions ($S$) leads to the mixing term $M_{NS}$. Finally we get the mass $M_S$ for sterile fermions from their $L$-violating interactions between the singlet sterile fermions. Consequently the neutrino mass matrix from eq. (\ref{secIIeq1}) takes the form:
	\begin{equation}
	\mathcal{M}= \begin{pmatrix}
	0&M^T_D&0\\M_D&0&M_{NS}\\0&M^T_{NS}&M_S
	\end{pmatrix}_{9\times9}
	\label{secIIeq2}
	\end{equation} 
	The matrix in eq. (\ref{secIIeq2}) is much useful in determining the $3\times3$ light neutrino mass matrix. Block diagonalisation of this mass matrix helps to obtain the desired structure of the matrix for the active neutrinos. For inverse seesaw the form of this important matrix turns out to be \cite{Malinsky:2009df}:
	\begin{equation}
	m_\nu=M^T_D(M^T_{NS})^{-1}M_SM^{-1}_{NS}M_D
	\label{secIIeq3}
	\end{equation}
	One can diagonalise this matrix in eq. (\ref{secIIeq3}), with the help of a unitary matrix, to get the mass eigenvalues of the three active neutrinos.

	\section{Model}{\label{secIII}}
	 In this work we have extended the mechanism of minimal inverse seesaw, ISS(2.3), with a Higgs-type scalar triplet $\eta=(\eta_1,\eta_2,\eta_3)$. This new addition plays a vital role in our study of neutrino phenomenology and aids in testing possible constraints from dark matter. Apart from this, we have used a few flavons which form important interactions with other fields of the model. These flavon fields are $ \phi $, $ \chi $, $ \chi^\prime$, $\zeta $ and $ \zeta^\prime$. In order to describe the relevant interactions among different particles we have considered the non-abelian discrete symmetry group $A_4$. For this purpose the particles of the model have specific assignments of the irreducible representation of the group. Accordingly, the lepton doublets ($L$), sterile fermions ($S_i$), scalar field $\eta$ and the flavons ($ \chi $, $ \chi^\prime$, $\zeta $, $ \zeta^\prime$) are considered to be triplets under the group $A_4$. The right-handed neutrinos ($N_1$, $N_2$) transform as $ 1^\prime $ and $ 1^{\prime\prime} $; whereas the Higgs field $H$ transforms as trivial singlet (1) under this group. Along with $A_4$ we have also used the abelian symmetry group $Z_3$ in this work. The assignment of charges to the particles have been highlighted in table (\ref{table1}).
	
	\begin{table}[ht]
			\footnotesize
			\begin{tabular}{|c| c c c c c c c c c c c|}
				\hline
				Fields & L &  $N_1$ & $N_2$ & $S_i$ & H & $\phi$ & $\eta$ & $\chi$ & $\chi^\prime$ & $\zeta$ & $\zeta^\prime$ \\
				\hline
				$A_4$ & 3 & $1^\prime$ & $1^{\prime\prime}$ & 3 & 1 & 1 & 3 & 3 & 3 & 3 & 3\\
				
				$Z_3$ & $\omega^2$ & $\omega$ & $\omega$ & 1 & $\omega$ & $\omega$ & $\omega^2$ & $\omega$ & $\omega$ & $\omega^2$ & $\omega^2$ \\
				\hline
			\end{tabular}
			\caption{Charge assignments of the particles under the various groups considered in the model.}
			\label{table1}
	\end{table}

    One of the aim of this work is to study the effect of dark matter in our setup. For this case the neutral components of $\eta$ are the probable dark matter candidates. We know that the discrete symmetry group $Z_2$ plays a crucial role in stabilizing the dark matter candidate.  Under $Z_2$ group, all the Standard Model particles are considered even and rest of the extra additional particles are taken to be odd. After electroweak symmetry breaking, only one component of $\eta$ acquires VEV and the other two components do not acquire any such value. At this state the $\eta$ fields can be expressed as \cite{Boucenna:2011tj}:
    \begin{equation}
    \eta_1=\begin{pmatrix}
    \eta_1^+\\ \frac{v_\eta+h_1+iA_1}{\sqrt{2}} 
    \end{pmatrix},	\hspace{5mm}	\eta_2=\begin{pmatrix}
    \eta_2^+\\ \frac{h_2+iA_2}{\sqrt{2}}
    \end{pmatrix}, \hspace{5mm} 	\eta_3=\begin{pmatrix}
    \eta_3^+\\ \frac{h_3+iA_3}{\sqrt{2}}
    \end{pmatrix}
    \label{sec3eq1}
    \end{equation}
    From eq. (\ref{sec3eq1}) we see that only one of the $\eta$'s acquire VEV i.e. $<\eta>=v_\eta(1,0,0)$ \cite{Hirsch:2010ru}. 
    Now the lagrangian for charged lepton sector can be written as:
    \begin{equation}
    L_L=a_1E^c_1H_d(L\phi)_1+ a_2 E^c_2H_d(L\phi)_{1'}+ a_3E^c_3H_d(L\phi)_{1''}
    \label{eqn:sec3eq2}
    \end{equation}
    For obtaining a diagonal charged lepton mass matrix, VEV of $\phi$ is considered as $\phi=(u,0,0)$. 
    So the mass matrix takes: $	M_L=diag(a_1,a_2,a_3)uv$. Moreover the parameters $a_1,a_2,a_3$ can be adjusted as per the need of the model. Finally the Yukawa lagrangian for the neutrinos can be written as:
	\begin{equation}
	\mathcal{L}= \frac{y_1}{\Lambda}N_1(L\chi)_3\eta +\frac{y_2}{\Lambda} N_2(L\chi^\prime)_3\eta + \gamma_1N_1(S\zeta)_{1''} + \gamma_2N_2(S\zeta^\prime)_{1'}+p(SS)_1
	\label{eq7}
	\end{equation}
	In eq.(\ref{eq7}) $y_1,y_2,\gamma_1,\gamma_2$ are the coupling constants. $\Lambda$ which is present in the above equation is the cut-off scale. The first two terms in the equation represents the Dirac mass term $M_D$. Third and the fourth terms are for mixing between right-handed neutrinos and sterile fermions $M_{NS}$. The final part denotes the mass term for sterile fermions ($M_S$). Following the $A_4$ multiplication rules, along with the VEV of $<\chi>=(\chi_1,\chi_2,\chi_3)$ and $\chi^\prime=(\chi_1^\prime,\chi_2^\prime,\chi_3^\prime)$, matrix for the Dirac mass can be written as:
	\begin{equation}
	M_D=\frac{v_\eta}{\Lambda}\begin{pmatrix}
	-\chi_3y_1 & -\chi^\prime_2y_2\\2\chi_2y_1 & -\chi^\prime_1y_2\\-\chi_1y_1 & 2\chi^\prime_3y_2
	\end{pmatrix}=\begin{pmatrix}
	a & b\\
	2c & d\\
	e & 2f
	\end{pmatrix}
	\label{sec3eq4}
	\end{equation}
	where $a=-\frac{v_\eta}{\Lambda}\chi_3y_1$, $b=-\frac{v_\eta}{\Lambda}\chi^\prime_2y_2$, $c=\frac{v_\eta}{\Lambda}\chi_2y_1$, $d=-\frac{v_\eta}{\Lambda}\chi^\prime_1y_2$, $e=-\frac{v_\eta}{\Lambda}\chi_1y_1$ and $f=\frac{v_\eta}{\Lambda}\chi^\prime_3y_2$.

	In a similar way, VEV alignment of the flavons $\zeta$ and $\zeta^\prime$ are considered to be $<\zeta>=v_\zeta(1,0,1)$ and $<\zeta^\prime>=v_{\zeta^\prime}(1,0,0)$. With the help of these alignments, we can express the matrices $M_{NS}$ and $M_S$ in the following way:

	 \begin{equation} 
	 M_{NS}=\begin{pmatrix}
	\gamma_1v_\zeta & 0 & \gamma_1v_\zeta\\0 & \gamma_2v_{\zeta^\prime} & 0
	\end{pmatrix}=\begin{pmatrix}
	g & 0 & g\\ 0 & h & 0
	\end{pmatrix},\qquad	
	M_S=p\begin{pmatrix}
	1 & 0 & 0\\0 & 0 & 1\\0 & 1 & 0
	\end{pmatrix}
	\label{sec3eq5}
\end{equation}
	
	In the above eq. (\ref{sec3eq5}) , $g=\gamma_1v_\zeta$ and $h=\gamma_2v_{\zeta^\prime}$. Now using these three matrices, we can construct the $8\times8$ neutrino mass matrix. For ISS(2,3) this matrix takes the following form:
	\begin{equation}
	\mathcal{M}= \begin{pmatrix}
	0&M^T_D&0\\M_D&0&M_{NS}\\0&M^T_{NS}&M_S
	\end{pmatrix}_{8\times8}
	\label{sec3eq6}
	\end{equation}
	This $8\times8$ matrix can be diagonalised with the help of a unitary matrix. The eight eigenvalues obtained after diagonalisation will correspond to the mass of the eight particles that are involved in the matrix. From eq. (\ref{sec3eq5}) it can be seen that $M_{NS}$ is a rectangular matrix. As a result, inverse of this matrix is not possible. Due to this issue, the expression for light neutrino mass matrix slightly changes to that in eq. (\ref{secIIeq3}). So in minimal inverse seesaw this expression is written as \cite{Abada:2014vea,Abada:2014zra}:
	\begin{equation}
	m_\nu=M_D.d.M^T_D
	\label{sec3eq7}
	\end{equation}
	In the above equation, $d$ is a $2\times 2$ matrix which can be derived from the $5\times 5$ heavy neutrino mass matrix $M_H$. The form of $d$ can be obtained in the following way \cite{Abada:2015rta}:
	\begin{equation}
	M_H^{-1}=\begin{pmatrix}
	0 & M_{NS} \\ M_{NS}^T & M_S
	\end{pmatrix}^{-1}=\begin{pmatrix}
	d_{2\times 2} & .....\\...... & .....
	\end{pmatrix}
	\label{sec3eq8}
	\end{equation}
	
	The matrix $m_\nu$ is symmetric in nature. Thus the elements of $m_\nu$ in terms of model parameters can be written as:
	\begin{equation}
	\begin{aligned}
	m11 &= -\frac{2abp}{gh}+\frac{b^2p}{h^2}\\
	m12 &=-\frac{bcp}{gh}+\frac{bdp}{h^2}-\frac{adp}{gh}\\
	m13 &=-\frac{bep}{gh}+\frac{bfp}{h^2}-\frac{afp}{gh}\\
	m22 &=-\frac{2cdp}{gh}+\frac{d^2p}{h^2}\\
	m23 &=-\frac{dep}{gh}+\frac{dfp}{h^2}-\frac{cfp}{gh}\\
	m33 &=-\frac{2efp}{gh}+\frac{f^2p}{h^2}\\
	\end{aligned}
	\end{equation}
	
	In this way, by using $A_4$ symmetry group, we have constructed a model in the mechanism of ISS(2,3). The eigenvalues of eq. (\ref{sec3eq7}) are the masses of three light neutrinos. Once this is done, it will be helpful to study other related phenomenologies.

	\section{Texture Zero Structures of the Matrices}{\label{secIV}}
	Texture zeros is an effective method that helps to reduce the number of free parameters associated with the light neutrino mass matrix. These texture structures are classified as one-zero, two-zero, three-zero etc. This classification is based on the number of elements of the mass matrix that are considered to be zero. However, only one-zero and two-zero textures are able to produce the neutrino parameters; the other texture structures are discarded as they are not compatible with the current experimental data. In this work we try to find out the impact on these texture structures by applying two-zero textures to Dirac mass matrix $M_D$ i.e. two elements of $M_D$ are simultaneously taken to be zero. We find that there are fifteen such possibilites of $M_D$. These are presented in the table (\ref{texturetab1}).
		\begin{table}[h]
		\begin{tabular}{|c|c|c|}
			\hline
			\multicolumn{3}{|c|}{Possible two zero textures of Dirac mass matrix $M_D$}\\
			\hline
			$M_{D1}=\begin{pmatrix}
			0 & 0\\ c & d\\ e & f\\
			\end{pmatrix}$ & $M_{D2}=\begin{pmatrix}
			0 & b\\ 0 & d\\ e & f\\
			\end{pmatrix}$ & $M_{D3}=\begin{pmatrix}
			0 & b\\ c & 0\\ e & f\\
			\end{pmatrix}$\\
			\hline
			$M_{D4}=\begin{pmatrix}
			0 & b\\ c & d\\ 0 & f\\
			\end{pmatrix}$ & $M_{D5}=\begin{pmatrix}
			0 & b\\ c & d\\ e & 0\\
			\end{pmatrix}$ & $M_{D6}=\begin{pmatrix}
			a & 0\\ 0 & d\\ e & f\\
			\end{pmatrix}$\\
			\hline
			$M_{D7}=\begin{pmatrix}
			a & 0\\ c & 0\\ e & f\\
			\end{pmatrix}$ & $M_{D8}=\begin{pmatrix}
			a & 0\\ c & d\\ 0 & f\\
			\end{pmatrix}$ & $M_{D9}=\begin{pmatrix}
			a & 0\\ c & d\\ e & 0\\
			\end{pmatrix}$\\
			\hline
			$M_{D10}=\begin{pmatrix}
			a & b\\ 0 & 0\\ e & f\\
			\end{pmatrix}$ & $M_{D11}=\begin{pmatrix}
			a & b\\ 0 & d\\ 0 & f\\
			\end{pmatrix}$ & $M_{D12}=\begin{pmatrix}
			a & b\\ 0 & d\\ e & 0\\
			\end{pmatrix}$\\
			\hline
			$M_{D13}=\begin{pmatrix}
			a & b\\ c & 0\\ 0 & f\\
			\end{pmatrix}$ & $M_{D14}=\begin{pmatrix}
			a & b\\ c & 0\\ e & 0\\
			\end{pmatrix}$ & $M_{D15}=\begin{pmatrix}
			a & b\\ c & d\\ 0 & 0\\
			\end{pmatrix}$\\
			\hline
		\end{tabular}
		\caption{The possible two zero textures of Dirac mass matrix $M_D$.}
		\label{texturetab1}
	\end{table} 
	
	It is interesting to note that out of all the fifteen possibilities, only six of them are able to produce one-zero textures of neutrino mass matrix. These structures are $M_{D3}$, $M_{D5}$, $M_{D6}$, $M_{D8}$, $M_{D12}$ and $M_{D13}$. They successfully produce diagonal one-zero textures of the mass matrix. Below we divide them into six different cases and try to explain the texture origin of neutrino mass matrix for our work.
	
	{\bf Case 1}: \\For the Dirac mass matrix of the form $M_{D3}$, the VEV of the flavons after symmetry breaking must be in the order:
	\begin{equation*}
		<\chi>=v_\chi(1,1,0)  \qquad <\chi^\prime>=v_{\chi^\prime}(0,1,1)
	\end{equation*}
	This particular VEV alignment gives 1-0 texture at (2,2) position,
	\begin{equation*}
		\begin{pmatrix}
		X & X & X\\ X & 0 & X\\ X & X & X\\
		\end{pmatrix}
	\end{equation*}
	
	{\bf Case 2}:\\
	For the matrix of the form $M_{D5}$, the required VEV alignments of the flavons are:
	\begin{equation*}
	<\chi>=v_\chi(1,1,0)  \qquad <\chi^\prime>=v_{\chi^\prime}(1,1,0)
	\end{equation*}
	This VEV arrangement of the flavons produces 1-0 textures at (3,3) position,
	\begin{equation*}
		\begin{pmatrix}
		X & X & X\\ X & X & X\\ X & X & 0\\
		\end{pmatrix}
	\end{equation*}
	
	{\bf Case 3}: \\For the Dirac mass matrix of the form $M_{D6}$, the VEV of the flavons after symmetry breaking must be in the order:
	\begin{equation*}
	<\chi>=v_\chi(1,0,1)  \qquad <\chi^\prime>=v_{\chi^\prime}(1,0,1)
	\end{equation*}
	This particular VEV alignment gives 1-0 texture at (1,1) position,
	\begin{equation*}
	\begin{pmatrix}
	0 & X & X\\ X & X & X\\ X & X & X\\
	\end{pmatrix}
	\end{equation*}
	
	{\bf Case 4}:\\
	For the matrix of the form $M_{D8}$, the required VEV alignments of the flavons are:
	\begin{equation*}
	<\chi>=v_\chi(0,1,1)  \qquad <\chi^\prime>=v_{\chi^\prime}(1,0,1)
	\end{equation*}
	This VEV arrangement of the flavons produces 1-0 textures at (1,1) position,
	\begin{equation*}
	\begin{pmatrix}
	0 & X & X\\ X & X & X\\ X & X & X\\
	\end{pmatrix}
	\end{equation*}
	
	{\bf Case 5}: \\For the Dirac mass matrix of the form $M_{D12}$, the VEV of the flavons after symmetry breaking must be in the order:
	\begin{equation*}
	<\chi>=v_\chi(1,0,1)  \qquad <\chi^\prime>=v_{\chi^\prime}(1,1,0)
	\end{equation*}
	This particular VEV alignment gives 1-0 texture at (3,3) position,
	\begin{equation*}
	\begin{pmatrix}
	X & X & X\\ X & X & X\\ X & X & 0\\
	\end{pmatrix}
	\end{equation*}
	
	{\bf Case 6}: \\For the Dirac mass matrix of the form $M_{D13}$, the VEV of the flavons after symmetry breaking must be in the order:
	\begin{equation*}
	<\chi>=v_\chi(0,1,1)  \qquad <\chi^\prime>=v_{\chi^\prime}(0,1,1)
	\end{equation*}
	This particular VEV alignment gives 1-0 texture at (2,2) position,
	\begin{equation*}
	\begin{pmatrix}
	X & X & X\\ X & 0 & X\\ X & X & X\\
	\end{pmatrix}
	\end{equation*}
	We perform our study on these six different cases. We try to find further constraints on these matrices based on our study of neutrino parameters, mixing angles and masses. This work is also extended to probe the effect of dark matter in this system.

	\section{Dark Matter}{\label{secVI}}
	Dark matter is one of the unsolved mysteries of the universe. Several attempts have been made to explore the nature of dark matter and find a suitable candidate for it. But as of now there is no unanimous agreement on a particle that can serve to be a suitable candidate. Physicists believe that in the early universe all the particles were in thermal equilibrium with each other. In the course of evolution, due to many factors this state of equilibrium was disturbed. As a result, density of some particle species became too low and their abundance in the universe became constant. This process is popularly known as freeze-out and the density of the particles at this stage is termed as relic density. In order to be a suitable dark matter candidate, a particle $\chi$ has to satisfy its relic density. The Boltzman equations for a dark matter candidate $\chi$ in terms of number density $n_\chi$ and density of $\chi$ when it was in thermal equilibrium ($n_\chi^{eqb}$), can be written in the following way \cite{Gondolo:1990dk,Griest:1990kh}:
	\begin{equation}
	\frac{dn_\chi}{dt}+3\mathcal{H}n_\chi=-\textless\sigma v\textgreater\left( n_\chi^2-\left( n_\chi^{eqb}\right)^2\right) 
	\label{sec6eq1}
	\end{equation}
	In the above eq. (\ref{sec6eq1})  $\textless\sigma v\textgreater$ is the thermally averaged annihilation cross-section of the dark matter candidate. This is expressed in terms of mass of the dark matter ($m_\chi$), mass of  the Majorana fermion ($m_\psi$) and the coupling of interaction between the fermions and dark matter in the model ($y$). 
    	\begin{equation}
    \textless\sigma v\textgreater = \frac{v^2 y^4 m_\chi^2}{48\pi(m_\chi^2+m_\psi^2)^2} 
    \label{sec6eq2}
    \end{equation}
    where $v$ is the relative velocity of the relic particles. At the time of freeze-out this value is 0.3$c$. The solution of eq. (\ref{sec6eq1}) gives the relic density of the dark matter particle. As a result, the expression for relic density is found to be \cite{Bai:2013iqa,Bell:2013wua}:
	\begin{equation}
	\Omega_\chi h^2=\frac{3\times 10^{-27} \textnormal{cm}^3 \textnormal{s}^{-1}}{\textless \sigma v\textgreater}
	\label{sec6eq3}
	\end{equation}
	In our work the neutral components of $\eta$ i.e. $\eta_2,\eta_3$ are the dark matter candidates. Using the above equations we have calculated the relic densities for both normal and inverted hierarchies. These results along with their descriptions are shown in the next section.

	\section{Numerical analysis and results}{\label{secVII}}
	In this section we present the results of our work. For numerical evaluation of different quantities of the model, we have used the latest 3$\sigma$ nu-fit values of neutrino oscillation parameters \cite{Esteban:2020cvm}. These values are highlighted in table \ref{secVIItabI}.
		
	\begin{table}[ht]
		\begin{tabular}{|c | c | c|}
			\hline
			Oscillation parameters & Normal Ordering & Inverted Ordering \\
			\hline
			$\textnormal{sin}^2\theta_{12}$ & [0.269,0.343] & [0.269,0.343]\\
			\hline
			$\textnormal{sin}^2\theta_{23}$ & [0.407,0.618] & [0.411,0.621]\\
			\hline
			$\textnormal{sin}^2\theta_{13}$ & [0.02034,0.02430] & [0.020530,0.02436] \\
			\hline
			$\Delta m^2_{21}/10^{-5} \textnormal{eV}^2$ & [6.28,8.04] & [6.82,8.04] \\
			\hline
			$\Delta m^2_{31}/10^{-3} \textnormal{eV}^2$ & [2.431,2.598] & [2.412,2.583]\\
			\hline
		\end{tabular}
		\caption{The latest $3 \sigma$ nu-fit values of oscillation parameters.}
		\label{secVIItabI}
	\end{table}
	In our model the charged lepton mass matrix is diagonal. Therefore the mixing matrix in this case is the unitary PMNS matrix, $U_{PMNS}$. Using this unitary mixing matrix, we diagonalise the light neutrino mass matrix of the model via the standard relation $m_\nu=U^T \textnormal{diag}(m_1,m_2,m_3)U$, where $m_1,m_2,m_3$ are the masses of the active neutrinos. These masses have different forms for normal and inverted hierarchies. For normal hierarchy these are expressed as: $\textnormal{diag}(0,\sqrt{m_1^2+\Delta m^2_{solar}},\sqrt{m_1^2+\Delta m^2_{atm}})$ and for inverted hierarchy it becomes $\textnormal{diag}(\sqrt{m_3^2+\Delta m^2_{atm}}),\sqrt{\Delta m_{atm}^2+\Delta m^2_{solar}},0)$ \cite{Nath:2016mts}. Through this process we can calculate the masses of neutrinos, their mixing angles and other parameters of the model. Additional constraints for these quantities come from solar and atmospheric mass squared differences. Once these values are obtained, they are then used to study the cosmological event of dark matter from the model. Furthermore, we have considered the values of $p$ in the range (10-20) KeV and VEV of $\eta$ is taken to be (1-10) GeV. The mixing angles can be calculated from the elements of mixing matrix. Thus, they can be expressed as \cite{Behera:2021eut}:
	\begin{align}
	\textnormal{sin}^2\theta_{13}=|U_{e3}|^2,   \hspace{6mm} \textnormal{sin}^2\theta_{23}=\frac{|U_{\mu 3}|^2}{1-|U_{e3}|^2}, \hspace{6mm} \textnormal{sin}^2\theta_{12}=\frac{|U_{e 2}|^2}{1-|U_{e3}|^2},
	\label{sec7eq1}
	\end{align}
	As mentioned earlier, only six of the two-zero structures of Dirac mass matrix are able to produce 1-0 texture conditions of light neutrino matrix.	We have performed our study in these six cases and have used the expressions in eq. (\ref{sec7eq1}) to find out the values of mixing angles for all these possible cases in both normal and inverted hierarcy. Also the parameters $g$ and $h$ are considered in the ranges [$10^5, 10^6$] GeV and [$10^3, 10^4$] GeV. Interestingly we found that only two of the 1-0 textures corresponding to $M_{D3}$ and $M_{D6}$ structures of Dirac matrix successfully produce all the three mixing angles in the allowed range for normal hierarchy. The other cases fail to produce one or two of these angles. Of the three mixing angles, two of them are possible in these cases, while the third is not possible and vice versa. This is true for both normal and inverted hierarchy. Because of this reason, we focus our study on normal hierarchy of the two cases that can generate all the mixing angles. In table (\ref{tab4}) we have highlighted the possibilities of obtaining the mixing angles in different 2-0 textures of Dirac mass matrix. In the following paragraphs we have presented the plots that describe the variations among different parameters of the model.

	\begin{table}[ht]
		\centering{
		\begin{tabular}{|c|c|ccc|ccc|}
			\hline
			\multirow{2}{*}{Dirac Mass} & \multirow{2}{*}{1-0 Texture} & \multicolumn{3}{|p{2cm}|}{Normal Hierarchy} & \multicolumn{3}{|p{2cm}|}{Inverted Hierarchy} \\
			\cline{3-8}
			& & $\theta_{12}$ & $\theta_{13}$ & $\theta_{23}$ & $\theta_{12}$ & $\theta_{13}$ & $\theta_{23}$ \\
			\hline
			$M_{D3}=\begin{pmatrix}
			0 & b\\ c & 0\\ e & f\\
			\end{pmatrix}$ & 
			$\begin{pmatrix}
			X & X & X\\ X & 0 & X\\ X & X & X\\
			\end{pmatrix}$
			& \ding{52} &  \ding{52} & \ding{52} & \ding{52} & \ding{56} & \ding{52} \\
			\hline
			$M_{D5}=\begin{pmatrix}
			0 & b\\ c & d\\ e & 0\\
			\end{pmatrix}$ & 
			$\begin{pmatrix}
			X & X & X\\ X & X & X\\ X & X & 0\\
			\end{pmatrix}$
			& \ding{56} & \ding{56} & \ding{52} & \ding{56} & \ding{56} & \ding{52} \\
			\hline
			$M_{D6}=\begin{pmatrix}
			a & 0\\ 0 & d\\ e & f\\
			\end{pmatrix}$ & 
			$\begin{pmatrix}
			0 & X & X\\ X & X & X\\ X & X & X\\
			\end{pmatrix}$
			& \ding{52} & \ding{52} & \ding{52} & \ding{56} & \ding{56} & \ding{56} \\
			\hline
			$M_{D8}=\begin{pmatrix}
			a & 0\\ c & d\\ 0 & f\\
			\end{pmatrix}$ & 
			$\begin{pmatrix}
			0 & X & X\\ X & X & X\\ X & X & X\\
			\end{pmatrix}$
			& \ding{56} & \ding{56} & \ding{52} & \ding{56} & \ding{56} & \ding{52} \\
			\hline
			$M_{D12}=\begin{pmatrix}
			a & b\\ 0 & d\\ e & 0\\
			\end{pmatrix}$ & 
			$\begin{pmatrix}
			X & X & X\\ X & X & X\\ X & X & 0\\
			\end{pmatrix}$
			& \ding{56} & \ding{52} & \ding{56} & \ding{56} & \ding{56} & \ding{56} \\
			\hline
			$M_{D13}=\begin{pmatrix}
			a & b\\ c & 0\\ 0 & f\\
			\end{pmatrix}$ & 
			$\begin{pmatrix}
			X & X & X\\ X & 0 & X\\ X & X & X\\
			\end{pmatrix}$
			& \ding{52} & \ding{56} & \ding{52} & \ding{52} & \ding{56} & \ding{52} \\
			\hline
		\end{tabular}
	}
	\caption{The table above shows the possibility of occurence of mixing angles for the six 2-0 textures of $M_D$.}
	\label{tab4}
	\end{table}

	\begin{figure}
		\includegraphics[scale=0.35]{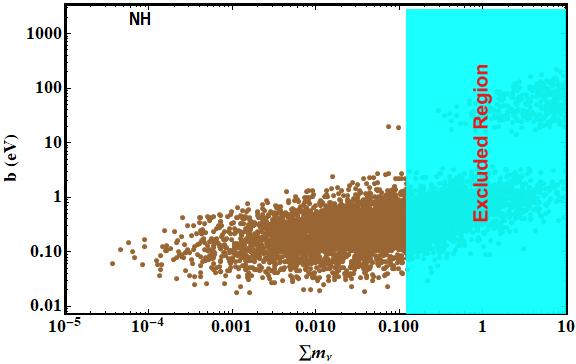}
		\includegraphics[scale=0.35]{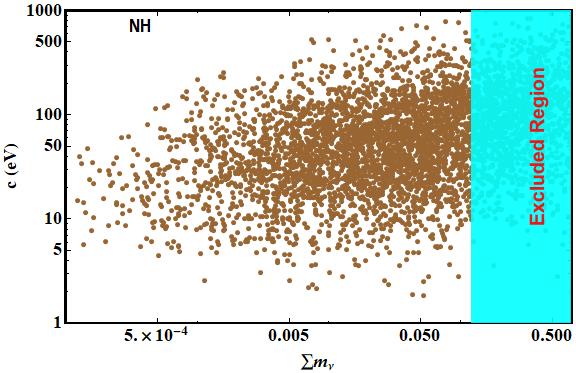}
		\includegraphics[scale=0.35]{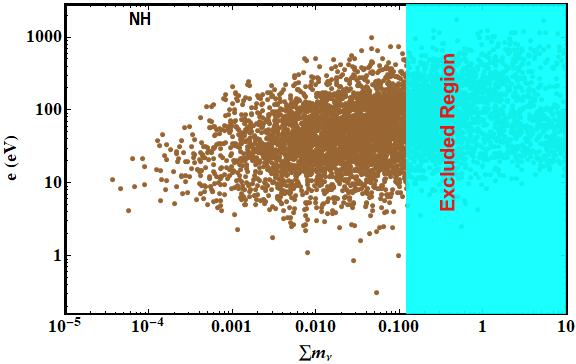}
		\includegraphics[scale=0.35]{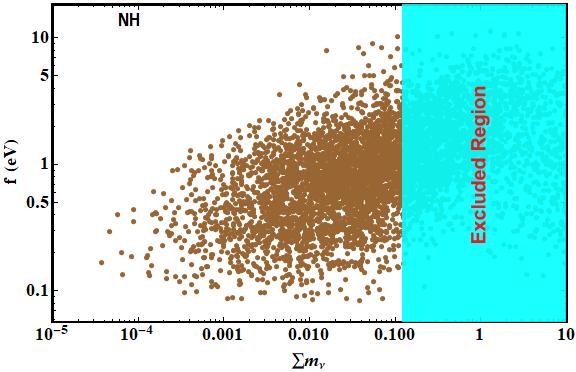}
		\caption{Corelation between the model parameters and $\sum m_\nu$ for Case 1 i.e. $M_{D3}$.}
		\label{fig1}
	\end{figure}
\begin{figure}
		\includegraphics[scale=0.35]{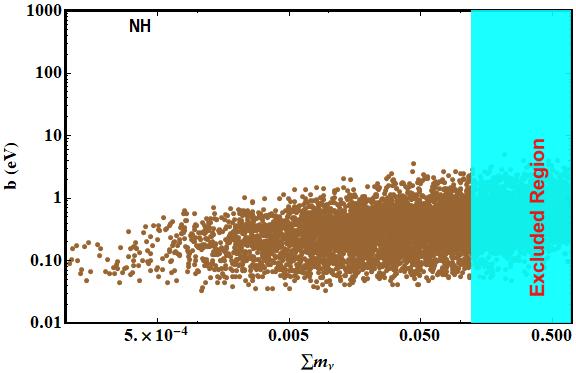}
		\includegraphics[scale=0.35]{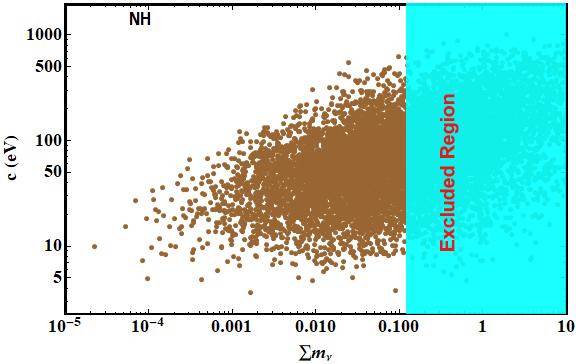}
		\includegraphics[scale=0.35]{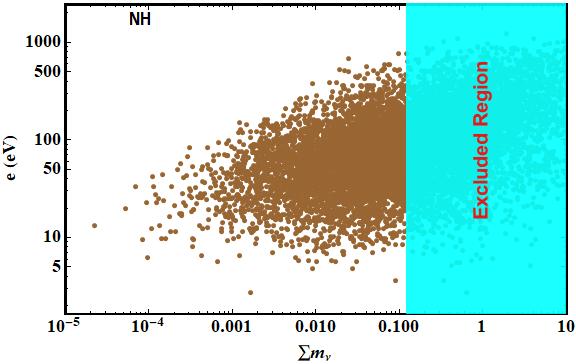}
		\includegraphics[scale=0.35]{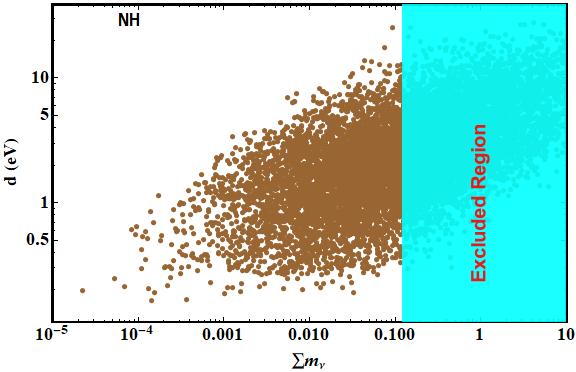}
		\caption{Corelation between the model parameters and $\sum m_\nu$ for Case 3 i.e. $M_{D6}$.}
		\label{fig2}
	\end{figure}

\newpage
  In fig. (\ref{fig1}) we have shown the variation of model parameters with respect to sum of neutrino mass ($\sum m_\nu$). From cosmological observations it is found that the upper limit for $\sum m_\nu$ is $\leq 0.12$ eV. From the figures we see that there are sufficient amounts of these parameters within the allowed range. For $b$ this range is found to be aroung (0.02-1) eV. The lower limit of the parameters $c$ and $e$ are almost same, around 5 eV. But the upper limit for $c$ is about 500 eV, whereas for $e$ it is around 900 eV. Also the range for $f$ is found to be (0.1-10) eV. Similarly the figures in (\ref{fig2}) show the relation between model parameters and $\sum m_\nu$ for \textit{Case 3 ($M_{D6}$)} of Dirac mass matrix. We find that $b$ lies in the range (0.05-1) eV. For this case, the parameters $c$ and $e$ lie in the same range (5-500) eV. Finally the values of $d$ are restricted in the region (0.2-10) eV. The common region of space for these parameters have been highlighted in table (\ref{tab5}).
  
  \begin{table}[h]
  	\begin{tabular}{|c|c|}
  		\hline
  		Model Parameters & Common Region (NH)\\
  		\hline
  		b  &  (0.05 - 1) eV \\
  		\hline
  		c  &  (5 - 500) eV\\
  		\hline
  		d  &  (0.2 - 10) eV \\
  		\hline
  		e  & (5 - 500) eV \\
  		\hline
  		f  &  (0.1 - 10) eV\\
  		\hline
  	\end{tabular}
  \caption{This table shows the favourable space of the model parameters.}
  \label{tab5}
  \end{table}

\begin{figure}
		\includegraphics[scale=0.35]{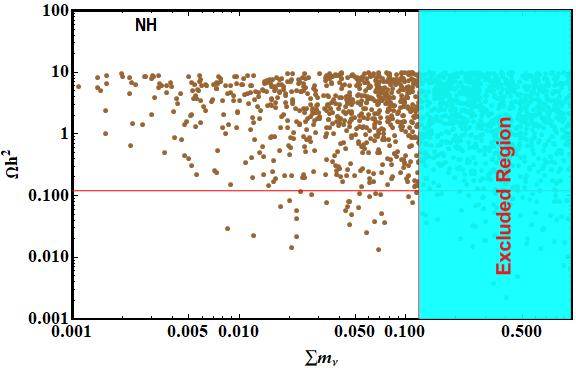}
		\includegraphics[scale=0.35]{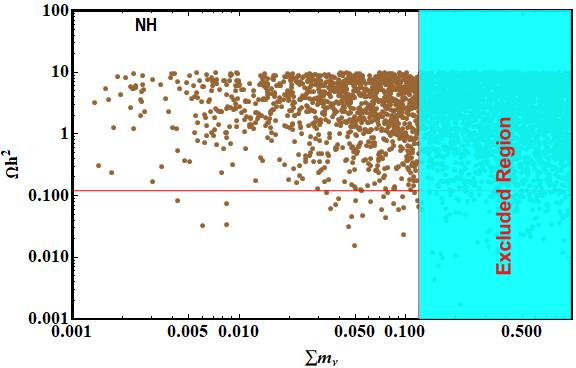}
		\caption{Corelation between relic density ($\Omega$h$^2$) and sum of neutrino mass ($\sum m_\nu$). The left (right) figure is for $M_{D3}$ ($M_{D6}$).}
		\label{fig3}
	\end{figure}

\begin{figure}
    	\includegraphics[scale=0.35]{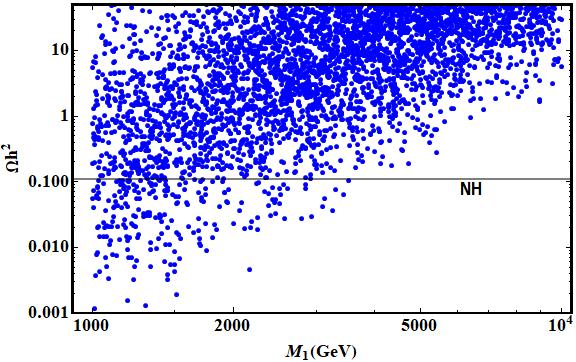}
    	\includegraphics[scale=0.35]{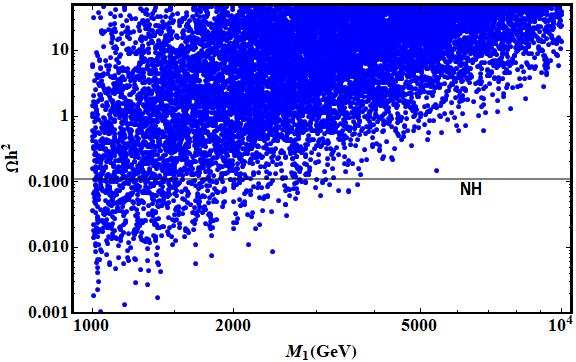}
    	\caption{Corelation between relic density ($\Omega$h$^2$) and right-handed neutrino $M_1$. The left (right) figure is for $M_{D3}$ ($M_{D6}$).}
    	\label{fig4}
    \end{figure}

	The plots in fig. (\ref{fig3}) and (\ref{fig4}) shows the relation between relic density of dark matter ($\Omega h^2$) with respect to sum of neutrino mass and lightest right-handed neutrino $M_1$ for both the 2-0 textures of Dirac mass matrix. From the figures it is clear that there are reasonable amounts of these quantities in the allowed range. In other words, the model is able to generate the relic density of dark matter for both the cases of $M_{D3}$ and $M_{D6}$. Also for the right-handed neutrino, the relic density is producible for a range of about (1000-5000) GeV of its mass. 
	
	\section{Conclusion}{\label{secVIII}}

	In this work we focus our study to realize the texture zero structures of neutrino mass matrix by implementing 2-0 conditions in Dirac mass matrix. For this purpose we have constructed a model in the framework of minimal inverse seesaw using the non-abelian discrete symmetry group $A_4$. There are five flavons present in this model $\chi, \chi^\prime, \zeta,\zeta^\prime, \phi$. We find that out of the fifteen possible 2-0 structures of $M_D$, only six of them are able to generate diagonal 1-0 texture structures of light neutrino mass matrix. Accordingly we take into account these six different cases and try to evaluate the model parameters for them. Furthermore we calculate the three neutrino mixing angles and masses to obtain any possible constraints on these six cases. Interestingly we find that only two of the structures of $M_D$ i.e. $M_{D3}$ and $M_{D6}$ favourably produce all the mixing angles in the allowed ranges for normal hierarchy. The other cases could generate two of the angles, but failed to produce the third angle and vice versa. As a result, we concentrate our study of dark matter in these two favourable cases and find that they are able to generate relic density in the desired range. So we can say that these results validate the compatibility of this model and it can be used to explore other phenomenological studies.

	\bibliography{file3}
	\bibliographystyle{utphys}

	\end{document}